\documentclass{article}

\usepackage{arxiv}
\usepackage{amsmath}
\usepackage[utf8]{inputenc} 
\usepackage[T1]{fontenc}    
\usepackage{hyperref}       
\usepackage{url}            
\usepackage{booktabs}       
\usepackage{amsfonts}       
\usepackage{nicefrac}       
\usepackage{microtype}      
\usepackage{lipsum}		
\usepackage{graphicx}
\usepackage{natbib}
\usepackage{doi}

\title{Bayesian Model Selection with Latent Group-Based Effects and Variances with the R Package slgf}

\author{ \href{https://orcid.org/0000-0003-3620-1405}{\includegraphics[scale=0.06]{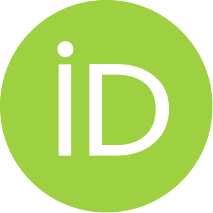}\hspace{1mm}Thomas A. ~Metzger}\\
	Department of Statistics\\
	The Ohio State University\\
	Columbus, OH 43210 \\
	\texttt{metzger.181@osu.edu} \\
	\And
	\href{https://orcid.org/0000-0003-1251-4378}{\includegraphics[scale=0.06]{orcid.pdf}\hspace{1mm}Christopher T. ~Franck} \\
	Department of Statistics\\
	Virginia Tech\\
	Blacksburg, Virginia \\
	\texttt{chfranck@vt.edu} \\
}

\renewcommand{\shorttitle}{\textit{arXiv} Bayesian Model Selection with slgf}

\hypersetup{
pdftitle={Bayesian Model Selection with Latent Group-Based Effects and Variances with the R Package slgf},
pdfsubject={stat},
pdfauthor={THomas A.~Metzger, Christopher T.~Franck},
pdfkeywords={model selection, Bayes factor, linear models},
}

\begin{document}
\maketitle

\begin{abstract}
 Linear modeling is ubiquitous, but performance can suffer when the model is misspecified. We have recently demonstrated that latent groupings in the levels of categorical predictors can complicate inference in a variety of fields including bioinformatics, agriculture, industry, engineering, and medicine. Here we present the R package slgf which enables the user to easily implement our recently-developed approach to detect group-based regression effects, latent interactions, and/or heteroscedastic error variance through Bayesian model selection. We focus on the scenario in which the levels of a categorical predictor exhibit two latent groups. We treat the detection of this grouping structure as an unsupervised learning problem by searching the space of possible groupings of factor levels. First we review the suspected latent grouping factor (SLGF) method. Next, using both observational and experimental data, we illustrate the usage of slgf in the context of several common linear model layouts: one-way analysis of variance (ANOVA), analysis of covariance (ANCOVA), a two-way replicated layout, and a two-way unreplicated layout. We have selected data that reveal the shortcomings of classical analyses to emphasize the advantage our method can provide when a latent grouping structure is present. 
\end{abstract}


\section{Introduction}\hypertarget{section:intro}{}

Linear models with categorical predictors (i.e., factors) are pervasive in the social, natural, and engineering sciences, among other fields. Conventional approaches to fit these models may fail to account for subtle latent structures, including latent regression effects, interactions, and heteroscedasticity within the data. These latent structures are frequently governed by the levels of a factor. Several examples of such datasets can be found in \citet{Franck2013}, \citet{FranckRJ}, \citet{KKSA}, and \citet{technometrics_paper}.  Our recent work \citep{technometrics_paper} developed latent grouping factor-based methodology to detect latent structures using Bayesian model selection. The current work provides an overview of the \textbf{slgf} package that enables users to easily implement the suspected latent grouping factor (SLGF) methodology, {and expands on the previous work by allowing for more flexible model specification.} 

\begin{figure}[htp!]
    \centering
    \includegraphics{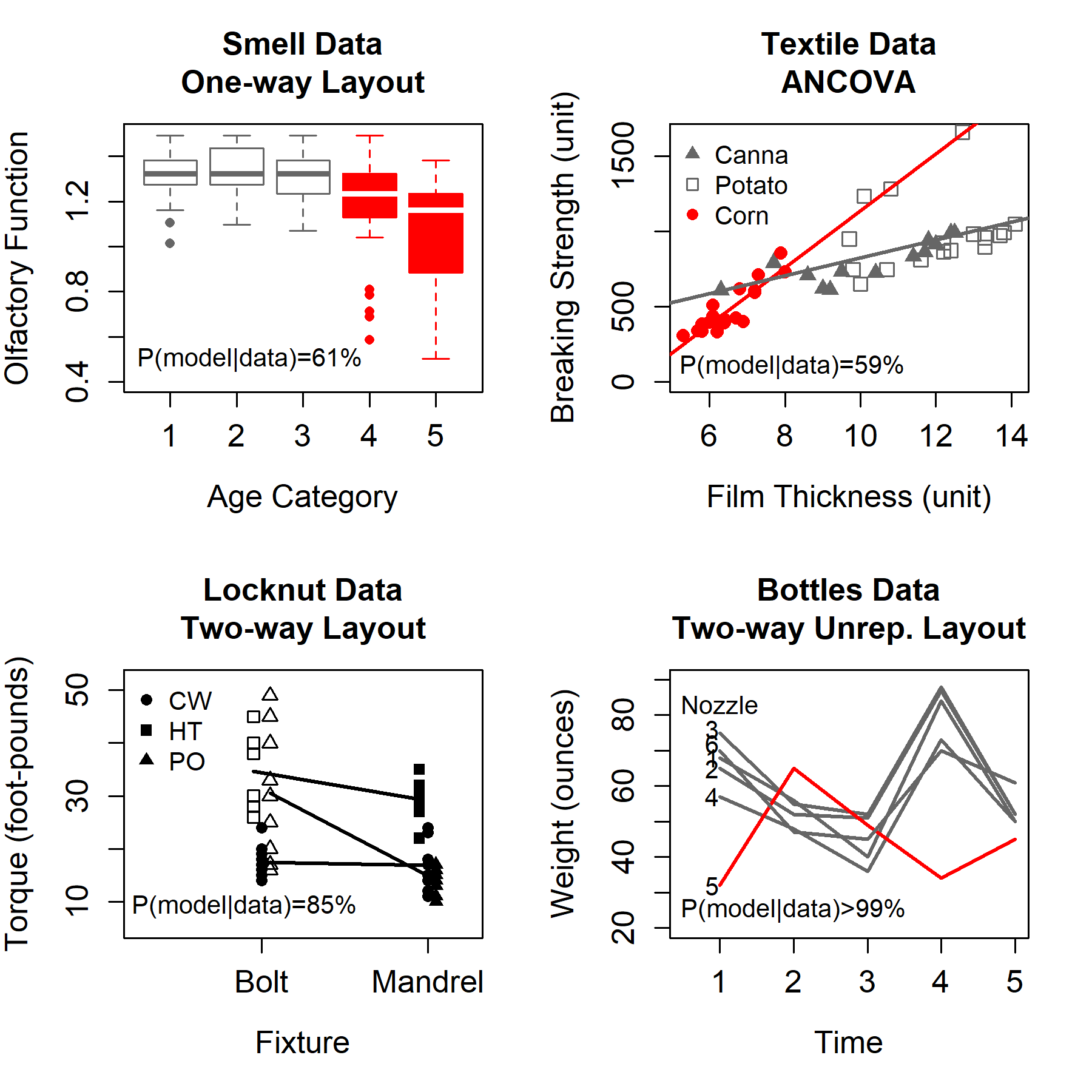}
    \caption{Smell data \citep[][top left]{smell}, textile data \citep[][top right]{Furry}, locknut data \citep[][bottom left]{locknut}, and bottles data \citep[][bottom right]{bottles}. Color (red/gray) shows latent grouping structure (i.e., group-based regression effects) for smell, textile, and bottles data, and fill (solid/open geometry) shows group-based variances for smell, textile, and locknut data.} 
    \label{figure:intro}
\end{figure}

Consider Figure \ref{figure:intro}, which illustrates four relevant data sets analyzed in this paper. In each panel, the levels of a user-specified factor are found to exhibit a latent grouping structure that partitions the data into two groups with distinct regression effects (indicated by color-coding) and/or error variances (filled and open geometry). With the \textbf{slgf} package, the user specifies the factor suspected of governing this latent structure. The package protects the user against detecting spurious latent grouping structures since it can accommodate non-grouped candidate models. It can also accommodate additional linear model terms of interest. The \textbf{slgf} package then assesses the plausibility of each model and the corresponding structures via Bayesian model selection. An overview of \textbf{slgf} functionality for these data follows and full details of each analysis (including candidate models) appear in Section \hyperlink{section:package}{\textbf{Using the slgf package}}. The \textbf{slgf} package focuses on assessing the plausibility of two-group structures in linear models with categorical predictors using fractional Bayes factors. A discussion comparing \textbf{slgf} and other R packages that address latent group models is in Section \hyperlink{section:conclusion}{\textbf{Conclusion}}.

The top left panel of Figure \ref{figure:intro} represents a one-way analysis of variance (ANOVA) study where a continuous measurement of olfactory function (vertical axis) is modeled as a function of age, where age is a factor represented in five categories (horizontal axis) \citep{smell}. We find the highest posterior model probability (61\%) for the model where levels 1, 2, and 3 of the SLGF age have distinct mean effects and error variances from levels 4 and 5. We call this the \texttt{smell} data set. 

The top right panel shows an analysis of covariance (ANCOVA), where the breaking strength of a starch film (vertical axis) is measured as a function of the SLGF (starch type) and a continuous measurement of film thickness (horizontal axis) \citep{Furry}. We find the highest posterior model probability (59\%) for the model where potato starch (unshaded gray squares) have a larger error variance than the shaded points, and, the red points (canna and corn starch) have a distinct slope from the gray points.  We call this the \texttt{textile} data set. 

The bottom left panel shows the example described by \citet{locknut}, where the torque required to tighten a locknut (vertical axis) was measured as a function of a plating process and a threading technique. The plating processes analyzed included treatments with cadmium and wax (CW), heat treating (HT), and phosphate and oil (PO). The threading techniques studied include bolt and mandrel, the types of fixture on which each locknut was affixed to conduct the test. We find the highest posterior model probability (85\%) for the model where bolt by HT and bolt by PO measurements have a larger error variance than those from bolt by CW, mandrel by HT, mandrel by PO, and mandrel by CW. We call this the \texttt{locknut} data. 

Finally, in the bottom right panel, the data set of \citet{bottles} represents an unreplicated two-way layout where six machine nozzles were used to fill bottles on five occasions (horizontal axis). The weight of each bottle (vertical axis) was measured, and we find the highest posterior model probability for the structure where nozzle 5 is found to be out of alignment from the others ($>99$\%). We call this the \texttt{bottles} data.

The \textbf{slgf} package implements a combinatoric approach that evaluates all possible assignments of SLGF levels into two groups. We refer to each these assignments as \textit{schemes}. For example, in the \texttt{smell} data, the scheme that is visualized assigns age levels 1, 2, and 3 into one group and levels 4 and 5 into the other, denoted \{1,2,3\}\{4,5\}. More details on how schemes are established can be found in Subsection \hyperlink{subsection:schemes}{Grouping schemes and model classes}.

The user may specify an SLGF for regression effects, another SLGF for error variances, require them to be the same, or specify no SLGF for one or both of these. For example, the \texttt{smell} data has age as the SLGF for both. In Subsection \hyperlink{subsection:textile}{\textbf{Case study 2: textile data}}, we analyze a data set with distinct regression and error variance SLGFs.  

In this paper, we provide an overview of the \textbf{slgf} package that enables analysis of data sets like those in Figure \ref{figure:intro} via Bayesian model selection. In Section \hyperlink{section:methodology}{\textbf{SLGF methodology}}, we briefly review the SLGF methodology. In Section \hyperlink{section:package}{\textbf{Using the slgf package}}, we illustrate the package functionality for the four data sets illustrated in Figure \ref{figure:intro}. For each data set, we will demonstrate the relevant code and package functionality along with a comparison between the results of a classical approach and our approach. In Section \hyperlink{section:conclusion}{\textbf{Conclusion}}, we summarize the package and its functionality. 

\section{SLGF methodology}\hypertarget{section:methodology}{}

\subsection{Model specification}\hypertarget{subsection:modelspec}{}

For a thorough review of the SLGF model specification see \citet{technometrics_paper}. First consider the linear model

\begin{equation}
\boldsymbol{Y}=\bm{1}^T \alpha + \bm{W}\boldsymbol{\nu}+\bm{V}\boldsymbol{\tau}+\bm{U}\boldsymbol{\rho}+\boldsymbol{\varepsilon,}\label{equation:WVUmodel}
\end{equation}

\noindent
where $1^T$ is an $N\times 1$ vector of 1s, $\alpha$ is an intercept, $\boldsymbol{\nu}$ represents the full SLGF effect with $K$ degrees of freedom, $\boldsymbol{\tau}$ represents the regression effects that do not arise from latent groupings (i.e., all other regression effects of interest), and the $\boldsymbol{\rho}$ terms indicate statistical interactions between SLGF and other regression effects; $W$, $V$, and $U$ partition the overall model matrix into model matrices corresponding to the SLGF effects $\boldsymbol{\rho}$, additional effects $\boldsymbol{\tau}$, and SLGF interactions, respectively; and finally $\boldsymbol{\varepsilon}$ represents an $N\times 1$ vector of errors where $\boldsymbol{\varepsilon}\overset{\text{iid}}{\sim}N(0,\Sigma)$ for $\Sigma=\sigma^2I$ where $I$ is an $N\times N$ identity matrix. 

Because a central goal of the SLGF methodology is to compare models with and without latent grouping structures, we next develop notation to indicate whether model terms in Equation (\ref{equation:WVUmodel}) involve groupings of factor levels or not. If a model contains a one degree of freedom group effect instead of the full $K$ degree of freedom SLGF effect, we denote the effect $\tilde{\boldsymbol{\nu}}$ instead, with corresponding $\tilde{W}$ to ensure they remain conformable. Similarly, if the interaction $\boldsymbol{\rho}$ is with the group effect rather than the full SLGF effect, we denote it $\tilde{\boldsymbol{\rho}}$. When there are group-based error variances, we let $\tilde{\boldsymbol{\varepsilon}}$ denote the vector of heteroscedastic errors, where the elements of $\tilde{\boldsymbol{\varepsilon}}$ are either $N(0,\sigma^2_1)$ or $N(0,\sigma^2_2)$ depending on their membership in group 1 or 2, respectively. 

For example, for the \texttt{smell} data in the top left panel of Figure \ref{figure:intro}, the most probable model can be represented as $Y=1^T\alpha + \tilde{W}\tilde{\boldsymbol{\nu}}+\tilde{\boldsymbol{\varepsilon}}$, with a 1 degree of freedom group effect $\tilde{\boldsymbol{\nu}}$ (color-coding) and heteroscedastic error term $\tilde{\boldsymbol{\varepsilon}}$ (shading). This model (posterior model probability 0.65) was found to be far more probable than the ordinary one way analysis of variance model $Y=1^T\alpha + W{\boldsymbol{\nu}}+{\boldsymbol{\varepsilon}}$ (posterior model probability less than 0.0001), the model with a 4 degree of freedom mean effect ${\boldsymbol{\nu}}$ and homoscedastic errors ${\boldsymbol{\varepsilon}}$. Similarly, the bottles data (bottom right panel) most probable model is $Y=1^T\alpha + W{\boldsymbol{\nu}}+\tilde{U}\tilde{\boldsymbol{\rho}}+{\boldsymbol{\varepsilon}}$ with a 4 degree of freedom nozzle effect ${\boldsymbol{\nu}}$, an 8 degree of freedom group-by-nozzle interaction $\tilde{\boldsymbol{\rho}}$, and homoscedastic errors $\boldsymbol{\varepsilon}$. 

\subsection{Grouping schemes and model classes}\hypertarget{subsection:schemes}
{}

Recall schemes are the possible assignments of factor levels to two latent groups. While the schemes shown in Figure \ref{figure:intro} may seem visually obvious, the \textbf{slgf} package considers all possible such assignments of factor levels into two groups. This (i) obviates the need for the user to specify specific schemes, and  (ii) apportions prior model probabilities commensurately with the actual number of models corresponding to a SLGF to prevent detection of spurious latent grouping structure. Problems will differ in the number of schemes under consideration. The package \textbf{slgf} automatically determines the schemes once the set of candidate models has been established by the user. The minimum number of levels that can comprise a grouping scheme can be adjusted by the user to lower the number of candidate models or to avoid creating model effects with too few degrees of freedom to be estimated. The user may specify the SLGF for regression effects and/or error variances, or neither. These SLGFs may or may not be different factors. If they are the same, the user may require that the grouping schemes must be equal or that they may be distinct. For example, in the textile data in the top right panel of Figure \ref{figure:intro}, the SLGF is starch for both regression effects and error variances, but the user should allow for distinct schemes since the variance scheme appears to be \{potato\}\{canna,corn\} and the regression effect scheme appears to be \{corn\}\{canna,potato\}. 

A model \textit{class} describes the structure of the model including specification of effects related to the hidden groups. Model classes may include, for example, the set of models with group-based regression effects but no group-based variances; or, a single model with no group-based regression effects or variances. For example, in the \texttt{smell} data represented in top left panel of Figure \ref{figure:intro}, we consider the following 62 models comprising six model classes: 

\begin{enumerate}
\item A single model with a 1 degree of freedom global mean effect and homoscedastic error variance; 
\item A single model with a 4 degree of freedom mean effect and homoscedastic error variance; 
\item 15 models (corresponding to the 15 possible grouping schemes) with a 1 degree of freedom global mean effect and group-based heteroscedastic error variances; 
\item 15 models with a 4 degree of freedom mean effect and group-based heteroscedastic error variances; 
\item 15 models with a 1 degree of freedom group-based mean effect and homoscedastic error variance; 
\item 15 models with a 1 degree of freedom group-based mean effect and group-based error variances. 
\end{enumerate}

For our analysis, we specified that the regression effect and variance grouping schemes must be equivalent, and that one level of the age factor could comprise a group. The user can relax these specifications as desired.  

\subsection{Parameter priors}\hypertarget{subsection:priors}{}
With \textbf{slgf}, the user can choose to implement noninformative priors on the regression effects (default), or the Zellner-Siow mixture of $g$-priors on these effects. We first enumerate the noninformative priors. Let $\boldsymbol{\beta}$ represent the full set of regression effects. For simplicity, we parametrize on the precision scale where $\varphi=\frac{1}{\sigma^2}$ and the corresponding precision matrix $\varphi I_{n \times n}$ is denoted $\Phi$. For a model $m_s^c$ where $c$ indexes class and $s$ indexes grouping scheme, \textbf{slgf} imposes 
\begin{equation}\label{eqn2}
P(\boldsymbol{\beta},\varphi|m_s^c)\propto \varphi
\end{equation}
for homoscedastic models, and 
\begin{equation}\label{eqn3}
P(\boldsymbol{\beta},\varphi_1, \varphi_2|m_s^c)\propto \varphi_1\cdot \varphi_2
\end{equation}

\noindent
for heteroscedastic models. 

Alternatively, in contexts with limited data, such as the two-way unreplicated \texttt{bottles} data in the bottom right panel of Figure \ref{figure:intro}, we recommend employing the Zellner-Siow mixture of $g$-prior \citep{Zellner1980, Zellner1986, Liangetal}, which reduces the minimal training sample size necessary for the computation of the fractional Bayes factor (see Subsection \hyperlink{subsection:FBF}{\textbf{Fractional Bayes factors and posterior model probabilities}} for further detail). We have generally found that in cases where the number of data points is close to the number of parameters in some of the larger candidate models (e.g., case study 4, bottles data), the mixture of $g$-priors approach outperforms the noninformative priors due to the drastic reduction in the required proportion of the data needed to implement the fractional Bayes factor approach. For homoscedastic models, recall $\Phi=\phi I$ where $I$ is an $N\times N$ identity matrix. Let 
\begin{equation}\label{eqn4}
P(\alpha,\varphi|m_s^c)\propto\varphi
\end{equation}
and 

\begin{equation}\label{eqn5}
\boldsymbol{\beta}_{-\alpha}|\Phi,g,m_s^c \sim N(\boldsymbol{0},\,g(X^T\Phi^{-1} X)^{-1}). 
\end{equation}

\noindent
Next, for heteroscedastic models, first denote $\tilde{\Phi}$ as a diagonal precision matrix where the $i$th diagonal element is either $\varphi_1$ or $\varphi_2$, depending upon the grouping membership of the $i$th observation. Let
\begin{equation}\label{eqn6}
P(\alpha, \varphi_1, \varphi_2|m_s^c)\propto \varphi_1\cdot\varphi_2
\end{equation}

\noindent
and 

\begin{equation}\label{eqn7}
\boldsymbol{\beta}_{-\alpha}|\tilde{\Phi},g,m_s^c \sim N(\boldsymbol{0},\,g(X^T\tilde{\Phi}^{-1} X)^{-1});
\end{equation}

\noindent 
In both homoscedastic and heteroscedastic cases, 
\begin{equation}\label{eqn8}
g\sim \text{InvGamma}\big(\frac{1}{2},\, \frac{N}{2}\big).
\end{equation}

\noindent 
Thus for homoscedastic models, the full prior on all parameters is the product of Equations (\ref{eqn4}), (\ref{eqn5}), and (\ref{eqn8}). For heteroscedastic models, it is the product of Equations (\ref{eqn6}), (\ref{eqn7}), and (\ref{eqn8}). 

\subsection{Fractional Bayes factors and posterior model probabilities}\hypertarget{subsection:FBF}{}

Note that if we form a standard Bayes factor for models using improper priors on parameters, the unspecified proportionality constants associated with the improper priors (Equations \ref{eqn2}, \ref{eqn3}, \ref{eqn4}, and \ref{eqn6}) would not cancel one another and the Bayes factor would be defined only up to an unspecified constant. Thus we invoke a fractional Bayes factor approach \citep{OHaganfbfs} to compute well-defined posterior model probabilities for each model. More details follow.

The \textbf{slgf} package obtains posterior model probabilities through the use of fractional Bayes factors. Briefly, a Bayes factor is defined as the ratio of two models' integrated likelihoods. The integrated likelihood is obtained by integrating parameters out of the joint distribution of data and parameters. In some cases, this integration is analytic, but in others, it is undertaken with a Laplace approximation; the corresponding simplified expressions and   methods used to optimize them are described in detail later in this section. In the SLGF context, let $\mathcal{M}$ represent the full set of models under consideration, representing all classes and grouping schemes of interest. Denote $\boldsymbol{\theta}$ as the full set of unknown parameters associated with a model $m_s^c\in \mathcal{M}$ and $\pi(\boldsymbol{\theta}|m_s^c)$ as the prior on these parameters given model $m_s^c$.  The parameter vector $\boldsymbol{\theta}$ depends on class and scheme of model $m_s^c$. The integrated likelihood is

$$P(\boldsymbol{Y}|m_s^c)=\int_{{\boldsymbol{\Theta}}} P(\boldsymbol{Y}|\boldsymbol{\theta},m_s^c)\pi(\boldsymbol{\theta}|m_s^c)d\boldsymbol{\theta},$$

\noindent
with Bayes factor comparing models $m_s^c$ and $m_{s'}^{c'}$

$$BF=\frac{P(\boldsymbol{Y}|m_s^c)}{P(\boldsymbol{Y}|m_{s'}^{c'})}.$$

Since the priors used by the \textbf{slgf} package are improper, $\pi(\boldsymbol{\theta}|m_s^c)$ is defined only up to an unspecified constant. Thus, $BF$ is defined only up to a ratio of unspecified constants. To overcome this issue and enable improper priors on parameters to be used in the course of Bayesian model selection, the fractional Bayes factor \citep{OHaganfbfs} was developed. A fractional Bayes factor is a ratio of two fractional marginal model likelihoods, where a fractional marginal likelihood is defined as
\begin{equation}\label{eqn:q}
q^b(\boldsymbol{Y}|m_s^c)=\frac{\int P(\boldsymbol{Y}|\boldsymbol{\theta},m_s^c)\pi(\boldsymbol{\theta}|m_s^c)d\boldsymbol{\theta}}{\int P(\boldsymbol{Y}|\boldsymbol{\theta},m_s^c)^b\pi(\boldsymbol{\theta}|m_s^c)d\boldsymbol{\theta}}.
\end{equation}

The $q^b(\boldsymbol{Y}|m_s^c)$ quantity in Equation (\ref{eqn:q}) is the integrated likelihood based on the $1-b$ fraction of the data where the improper prior has been updated to become proper with $b$ fraction of the data. Thus all normalizing constants are specified. The fractional Bayes factor is thus

$$FBF=\frac{q^b(\boldsymbol{Y}|m_s^c)}{q^b(\boldsymbol{Y}|m_{s'}^{c'})}.$$

\noindent 
for some fractional exponent $0<b<1$. Thus we must compute the integrals $\int P(\boldsymbol{Y}|\boldsymbol{\theta},m_s^c)\pi(\boldsymbol{\theta}|m_s^c)d\boldsymbol{\theta}$ and $\int P(\boldsymbol{Y}|\boldsymbol{\theta},m_s^c)^b\pi(\boldsymbol{\theta}|m_s^c)d\boldsymbol{\theta}$, the numerator and denominator of Equation (\ref{eqn:q}), respectively, for all $m_s^c\in \mathcal{M}$. Although \citet{OHaganfbfs} provides several recommendations for choice of $b$, \textbf{slgf} exclusively implements $b=\frac{m_0}{N}$ where $m_0$ is the minimal training sample size required for the denominator of Equation (\ref{eqn:q}) to be proper for all models. If $m_0$ is too small, then the denominator of Equation (\ref{eqn:q}) diverges. The user must specify $m_0$; if their choice is too low, then \textbf{slgf} increases it until all relevant integrals converge. For further details, see \citet{OHaganfbfs}, p. 101; for recommendations on choosing $m_0$ in practice, see Subsection \hyperlink{subsection:m0}{\textbf{Choice of }$\mathbf{m_0}$}. 

Next we discuss the technical details on how these integrals are computed via Laplace approximation. Specifically, we will describe how $\log\left(q^b(\boldsymbol{Y}|m_s^c)\right)$ is computed in each case. In the case of noninformative regression priors for homoscedastic models, $\boldsymbol{\beta}$ and $\sigma^2$ are integrated analytically. Let $\hat{\boldsymbol{Y}}$ represent the fitted values of $m_s^c$ and $\text{SSResid}$ the residual sum of squares of this model. We obtain 
\begin{equation}
\log\left(q^b(\boldsymbol{Y}|m_s^c)\right)=\left(-\frac{N(1-b)}{2}\right)(\log\pi+\log(\text{SSResid}))+\left({\frac{Nb-1}{2}}\right)\log b+\log\left(\frac{\Gamma\left(\frac{ N-P}{2}\right)}{\Gamma\left(\frac{Nb-P}{2}\right)}\right)
\end{equation}

\noindent In the case of noninformative regression priors for heteroscedastic models, both the numerator and denominator integrals of Equation (\ref{eqn:q} )must be approximated with a Laplace approximation because although $\boldsymbol{\beta}$ can be integrated analytically, $\sigma^2_1$ and $\sigma^2_2$ cannot be. The integrals are computed on the log-scale for numeric stability. Equation (\ref{eqn:q}) on the log-scale simplifies to:

\begin{equation}
\log\left(q^b(\boldsymbol{Y}|m_s^c)\right)=\frac{N(b-1)}{2}\log(2\pi)+\frac{P+1}{2}\log b + \frac{1}{2}\log\left(\frac{|H_b^{\star}|}{|H^{{\star}}|}\right) + \log\left(\frac{P(\boldsymbol{Y}|\boldsymbol{\theta}^{{\star}})\pi(\boldsymbol{\theta}^{\star}|m_s^c)}{P(\boldsymbol{Y}|\boldsymbol{\theta}_b^{{\star}})^b\pi(\boldsymbol{\theta}^{\star}_b|m_s^c)}\right)
\end{equation}

\noindent where $\boldsymbol{\theta}^{\star}$ and $H^{\star}$ denote the mode and Hessian of $P(\boldsymbol{Y}|\boldsymbol{{\theta}},m_s^c)\pi(\boldsymbol{\theta}|m_s^c)$, and $\boldsymbol{\theta}^{\star}_b$ and $H^{\star}_b$ denote the mode and Hessian of $P(\boldsymbol{Y}|\boldsymbol{{\theta}},m_s^c)^b\pi(\boldsymbol{\theta}|m_s^c)$. These modes and Hessians are computed with \textbf{optim} using the Nelder-Mead algorithm. 

In the Zellner-Siow mixture of  $g$-prior case, $\alpha$ and $\boldsymbol{\beta}_{-\alpha}$ are integrated analytically. For homoscedastic models, $\sigma^2$ is as well, and only $g$ is integrated with a Laplace approximation. Again marginal model likelihoods are computed on the log-scale. The log of the mode of $P(\boldsymbol{Y}|g,m_s^c)^b\pi(\boldsymbol{\theta}|m_s^c)$, denoted $g^{\star}_b$, is found by solving the closed-form equation $\frac{(Nb-1-P)}{2}\log(1+bg) + \frac{Nb-1}{2}\log(1+bg(1-R^2))-\frac{3}{2}\log g-\frac{N}{2g}:=0$ with the base \textbf{R} function \textbf{uniroot} where $R^2$ is the coefficient of determination for $m_s^c$. The Hessian is then evaluated at this solution $g^{\star}_b$; the closed-form Hessian of $P(\boldsymbol{Y}|g,m_s^c)^b\pi(\boldsymbol{\theta}|m_s^c)$ evaluated at $g^{\star}$ is given by $H^{\star}_b=\frac{1}{2}\left(\frac{((Nb-1)b^2(1-R^2)^2}{(1+bg^{\star})(1-R^2)^2}-\frac{(Nb-P-1)b^2}{(1+bg^{\star})^2}+\frac{3}{g^{{\star}^2}} - \frac{2N}{g^{{\star}^3}} \right)$. For $b=1$, this expression describes the numerator of Equation (\ref{eqn:q}); see \citet{Liangetal} for further mathematical details. The Laplace approximation for Equation (\ref{eqn:q}) on the log-scale then is given by:

\begin{equation}
\begin{split}
\log\left(q^b(\boldsymbol{Y}|m_s^c)\right)=&\log\left(\frac{\Gamma\left(\frac{N-1}{2}\right)}{\Gamma\left(\frac{Nb-1}{2}\right)}\right)+\frac{Nb-1}{2}\left(\log(\text{SSTotal})+\log\pi\right)+ \frac{1}{2}\log\left(\frac{|H_b^{\star}|}{|H^{{\star}}|}\right) + \\
& \ \ \log\left(\frac{P(\boldsymbol{Y}|\boldsymbol{\theta}^{{\star}})\pi(\boldsymbol{\theta}^{\star}|m_s^c)}{P(\boldsymbol{Y}|\boldsymbol{\theta}_b^{{\star}})^b\pi(\boldsymbol{\theta}^{\star}_b|m_s^c)}\right).
\end{split}
\end{equation}

For heteroscedastic models, a three-dimensional Laplace approximation is used to integrate $\sigma^2_1$, $\sigma^2_2$, and $g$. To obtain $\boldsymbol{\theta}^{\star}_b$ and $\boldsymbol{\theta}^{\star}$, we first transform $\gamma_1=\log\left(\frac{1}{\sigma^2_1}\right)$ and $\gamma_2=\log\left(\frac{1}{\sigma^2_2}\right)$ to stabilize the optimization. We optimize $\log P(\boldsymbol{Y}|g,\sigma^2_1,\sigma^2_2)^b\pi(\sigma^2_1,\sigma^2_2,g)=\frac{n_1b}{2}\log \gamma_1+\frac{n_2b}{2}\gamma_2-\frac{P}{2}\log g+\frac{1}{2}|X^T\tilde{\Sigma}X|-\frac{1}{2}\log|\frac{bg+1}{bg}X^T(\tilde{\Sigma}-Z_{\tilde{\Sigma}})X|-\frac{b}{2}\boldsymbol{Y}^T \left(\tilde{\Sigma} -Z_{\tilde{\Sigma}}-(\tilde{\Sigma}-Z_{\tilde{\Sigma}})\right)X \left(\frac{bg+1}{bg}X^T\tilde{\Sigma}X-X^TZ_{\tilde{\Sigma}}X^{-1}X^T(\tilde{\Sigma}-Z_{\tilde{\Sigma}})\right)\boldsymbol{Y}-\frac{3}{2}\log(g)-\frac{N}{2g}+\log(J)$ using the Nelder-Mead method from \textbf{optim} where  $Z_{\tilde{\Sigma}}=\tilde{\Sigma}Z(Z^T \tilde{\Sigma} Z)^{-1}Z^T \tilde{\Sigma}$, $Z=\boldsymbol{1}^T$, and $\log(J)=-(\gamma_1+\gamma_2)$ represents the determinant of the log-precision transformation. For $b=1$ these equations yield integrand of the numerator of (\ref{eqn:q}). 

With the modes computed, the Hessians of $\log P(\boldsymbol{Y}|g,\sigma^2_1,\sigma^2_2)^b\pi(\sigma^2_1,\sigma^2_2,g)$ are calculated with the function \textbf{Hessian} from the package \textbf{numDeriv}. Finally with the modes and Hessians computed, the Laplace approximation for Equation (\ref{eqn:q}) is given by:

\begin{equation}
\log \left(q^b(\boldsymbol{Y}|m_s^c)\right)=\frac{Nb-1}{2}\log(2\pi)+\frac{P+1}{2}\log(b)+ \frac{1}{2}\log\left(\frac{|H_b^{\star}|}{|H^{{\star}}|}\right) + \log\left(\frac{P(\boldsymbol{Y}|\boldsymbol{\theta}^{{\star}})\pi(\boldsymbol{\theta}^{\star}|m_s^c)}{P(\boldsymbol{Y}|\boldsymbol{\theta}_b^{{\star}})^b\pi(\boldsymbol{\theta}^{\star}_b|m_s^c)}\right).
\end{equation}

For the sake of consistency, all models, even with fully tractable marginal model likelihoods, are computed with a FBF. Once log-fractional marginal likelihoods have been computed for all models, we subtract the maximum from this set so that the set of log-fractional marginal likelihoods has been rescaled to have a maximum of 0. Each value is exponentiated to obtain a set of fractional marginal likelihoods with maximum 1. This adjustment helps to avoid numerical underflow when computing posterior model probabilities.

\subsection{Choice of minimal training sample size}\hypertarget{subsection:m0}{}

The user must specify the argument $m_0$, the minimal training sample size such that all marginal model likelihoods are well-defined. If \texttt{prior="flat"}, then we recommend that the user begins by letting \texttt{m0} equal the dimension of the improper priors: that is, the number of coefficients in most complex model under consideration plus the number of variances under consideration. If \texttt{prior="zs"}, then \texttt{m0} can generally be much smaller (in practice, we have found that \texttt{m0=2} performs well) as the prior on the regression effects is proper. If the user's choice is too low, then \texttt{ms\_slgf} will incrementally increase it by 1 until all marginal model probabilities are numerically stable. If \texttt{m0} reaches $n$, corresponding to 100\% of data used for training, \texttt{ms\_slgf} will terminate and the user should specify a different set of models. 

\subsection{Model priors}\hypertarget{subsection:modelpriors}{}

With this adjusted set of fractional marginal likelihoods, we next consider the priors for the model space. The function \texttt{ms\_slgf} imposes a uniform prior by model class, and for classes containing multiple models, the prior on each class is uniformly divided among the models it contains. We finally compute posterior model probabilities for each model: 

\begin{equation}
P(m^{\prime}|\boldsymbol{Y})=\frac{P(\boldsymbol{Y}|m^{\prime})P(m^{\prime})}{\underset{\mathcal{M}}{\sum}P(\boldsymbol{Y}|m)P(m)}.
\end{equation}

The prior probability placed on each model can be found in the \texttt{models\$ModPrior} vector in output from \texttt{ms\_slgf}. 

\subsection{Parameter estimation}\hypertarget{subsection:estimation}{}

Our approach provides maximum \textit{a posteriori} (MAP) estimates for all relevant quantities: $\hat{\boldsymbol{\beta}}$, $\hat{\boldsymbol{\sigma}}^2=\{\hat{\sigma}^2\}$ or $\hat{\boldsymbol{\sigma}}^2=\{\hat{\sigma}^2_1,\hat{\sigma}^2_2\}$ in the homoscedasitc and heteroscedastic cases respectively, and $g$ in the Zellner-Siow mixture of $g$-prior case. 

Because the prior on $\boldsymbol{\beta}$ is either flat or centered at 0, the MAP estimator is simply the usual maximum likelihood estimator:

\begin{equation}
\hat{\boldsymbol{\beta}}=\underset{\boldsymbol{\beta}}{\arg\max}P(\boldsymbol{Y}|X,\boldsymbol{\beta},\Sigma) 
\end{equation}
so that $\hat{\boldsymbol{\beta}}=(X^TX)^{-1}X^T\boldsymbol{Y}$. The variance(s) and $g$ were computed via the base \textbf{R} function \textbf{optim} during the Laplace approximation stage. For computational efficiency, ${\boldsymbol{\beta}}$ is integrated out of $P(\boldsymbol{Y}|X,\boldsymbol{\theta})P(\boldsymbol{\theta})$ and the variances are estimated on the log-scale, so we let $\hat{\boldsymbol{\lambda}}:=\{\hat{\lambda}\}$ in homoscedastic models or $\{\hat{\lambda}_1,\hat{\lambda}_2\}$ in heteroscedastic models. Then

\begin{equation}
\hat{\boldsymbol{\lambda}}=\underset{\boldsymbol{\lambda}}{\arg \max}\int P(\mathbf{Y}|X,\boldsymbol{\beta},\Sigma)P(\boldsymbol{\beta})P(\Sigma)d\boldsymbol{\beta}
\end{equation}
or, 
\begin{equation}
\{\hat{\boldsymbol{\lambda}},\hat{g}\}=\underset{\boldsymbol{\lambda},g}{\arg\max}\int P(\mathbf{Y}|X,\boldsymbol{\beta},\Sigma,g)P(\alpha)P(\boldsymbol{\beta}_{-\alpha}|\Sigma)P(g)d\boldsymbol{\beta}.
\end{equation}
Then, $\hat{\boldsymbol{\sigma}}^2=\exp\{\hat{\boldsymbol{\lambda}}\}$ for $\hat{\boldsymbol{\sigma}}^2=\{\hat{\sigma}^2\}$ or $\hat{\boldsymbol{\sigma}}^2=\{\hat{\sigma}^2_1,\hat{\sigma}^2_2\}$. The output values \texttt{coefficients}, \texttt{variances}, and \texttt{gs} (only if \texttt{prior="zs"}) are lists where each element contains the estimates for each model's  $\hat{\boldsymbol{\beta}}$, $\hat{\boldsymbol{\sigma}^2}$, and $\hat{\textbf{g}}$, respectively. 

\section{Using the slgf package}\hypertarget{section:package}{}

The function \texttt{ms\_slgf()} is the main function of \textbf{slgf} that implements the methodology we have described. Each argument of \texttt{ms\_slgf()} and its output will be illustrated in the case studies found in Subsections \hyperlink{subsection:smell}{\textbf{Case study 1: smell data}}, \hyperlink{subsection:textile}{\textbf{Case study 2: textile data}}, \hyperlink{subsection:torque}{\textbf{Case study 3: locknut data}}, and \hyperlink{subsection:bottles}{\textbf{Case study 4: bottles data}}. The \texttt{ms\_slgf()} function requires several inputs to compute and output posterior model probabilities for all models, schemes, and model classes of interest. The user begins with a \texttt{data.frame} containing a continuous response, at least one categorical predictor, and any other covariates of interest. The \texttt{data.frame} cannot contain column names with the character string \texttt{group}, because \texttt{ms\_slgf()} will search for this string when fitting group-based models. The user must first identify an SLGF for the fixed effects and/or the variance. The user indicates, via the arguments \texttt{response}, \texttt{slgf\_beta}, and \texttt{slgf\_Sigma}, character strings corresponding to the response, the suspected latent fixed effect grouping factor, and the suspected latent variance grouping factors, respectively. If no latent regression effect structure or variance structure is to be considered, the user may specify \texttt{slgf\_beta=NA}, \texttt{slgf\_Sigma=NA}, or both. We note that if the user does not specify any SLGFs, the model selection is still undertaken through fractional Bayes factors as described previously. If the user chooses the same categorical variable for both latent grouping factors, the argument \texttt{same\_scheme}, which defaults to \texttt{FALSE}, can indicate whether the grouping schemes for the regression effect and variance structures must be equivalent. 

Next the user determines the model classes they wish to evaluate. The argument \texttt{usermodels} is a \texttt{list} where each element contains a string of R class \texttt{formula} or \texttt{character}. The user also specifies which classes should also be considered in a heteroscedastic context via the argument \texttt{het}, which is a vector of the same length as \texttt{usermodels}, containing an indicator 1 or 0 corresponding to each model class specified in \texttt{usermodels} where 1 indicates the model will be considered with group-based variances and 0 indicates it will not. Together the arguments \texttt{usermodels} and \texttt{het} indicate which fixed effect structures are of interest, and which should be further considered for heteroscedasticity, thus implicitly creating the full set of model classes considered.  

Next the user chooses a prior to place on the regression effects. As described in Subsection \hyperlink{subsection:priors}{\textbf{Parameter priors}}, \texttt{prior="flat"} (the default) implements the noninformative prior and \texttt{prior="zs"} imposes the Zellner-Siow mixture of $g$-prior.  

Finally the user must specify the minimum number of levels of the SLGF that can comprise a group, via the arguments \texttt{min\_levels\_beta} and \texttt{min\_levels\_Sigma}, which default to 1. The number of possible grouping schemes increases with the number of levels of the SLGF. To speed up the computation, the user can increase these arguments and thus reduce the number of candidate models. Because we partition into two groups, note these arguments may not exceed half the number of levels of the SLGF. Additionally, when considering data with limited degrees of freedom, increasing \texttt{min\_levels\_beta} and/or \texttt{min\_levels\_Sigma} may be necessary to ensure effects can be computed. 

\subsection{Case Study 1: smell data}\hypertarget{subsection:smell}{}

\begin{figure}
    \centering
    \includegraphics{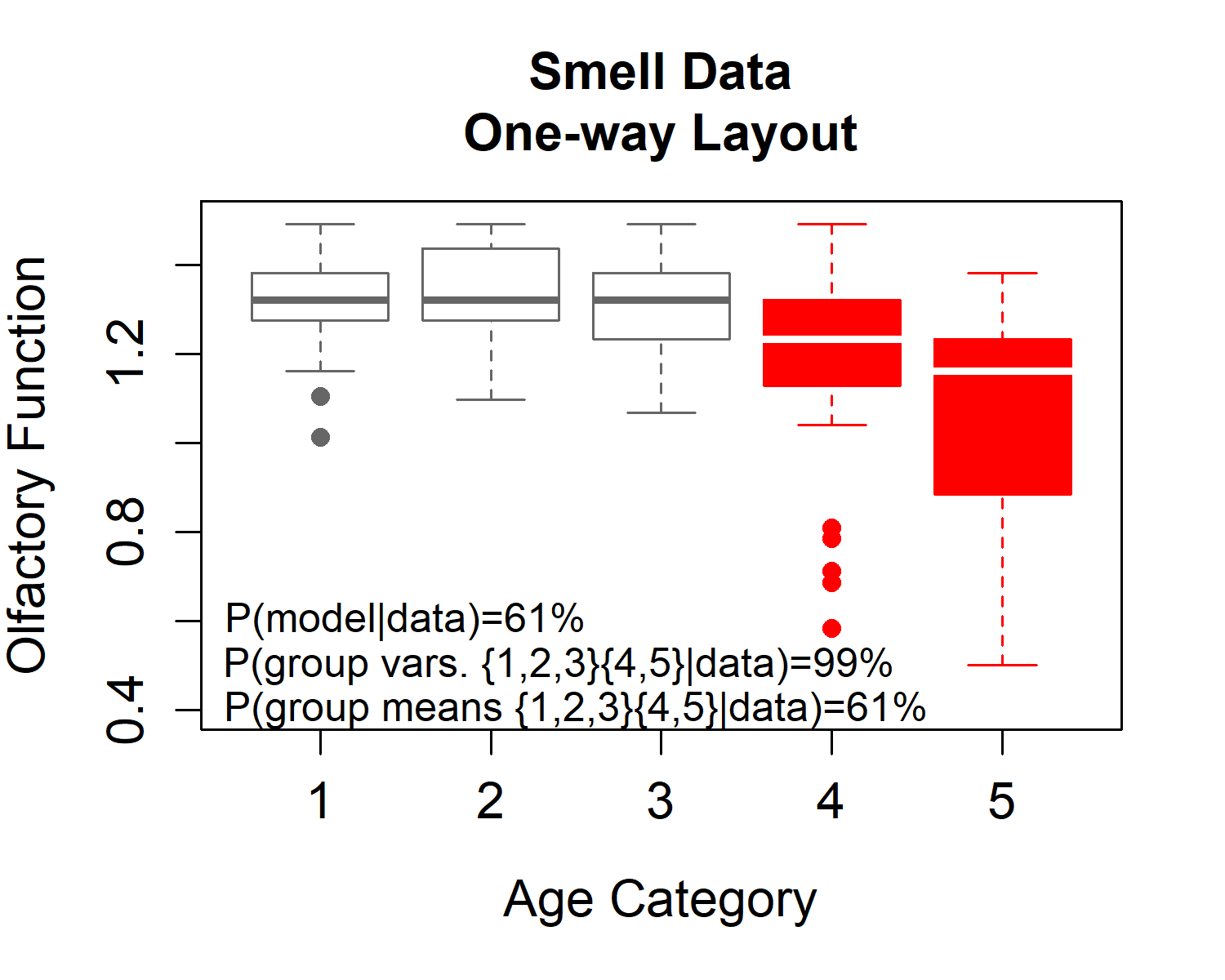}
    \caption{The smell data \citep{smell} is analyzed for group-based means and variances. We find posterior model probability of 61\% for the model with group-based means and variances with scheme \{1,2,3\}\{4,5\}. We also find overall posterior probability of grouping scheme }. 
    \label{figure:smell}
\end{figure}

First we revisit the smell data set analyzed by \citet{smell}. They measured olfactory acuity (denoted \texttt{olf}) on a continuous scale as a function of age (\texttt{agecat}), where age groups were divided into five categorical levels. See Figure \ref{figure:smell}. We note that levels 4 and 5 of \texttt{agecat} appear to have larger variance than levels 1, 2, and 3, but standard analysis of variance models assume homoscedasticity. We first demonstrate how a classical analysis might misrepresent the data. A usual one-way ANOVA analysis compares the null model, with a single mean, against the alternative model, with 4 degrees of freedom for the mean effects, with homoscedastic error variance. The user may need to coerce agecat to a factor variable. 

\begin{example} 

> smell_null <- lm(olf~1, data=smell)      # fit a null model with a single mean
> smell_full <- lm(olf~agecat, data=smell) # fit a full model with a 4 agecat effects
> print(smell_null) 
Call:
lm(formula = olf ~ 1, data = smell)

Coefficients:
(Intercept)  
      1.234 
> print(smell_full) 
Call:
lm(formula = olf ~ agecat, data = smell)

Coefficients:
(Intercept)      agecat2      agecat3      agecat4      agecat5  
    1.31689      0.02824     -0.01075     -0.11580     -0.25728  
> anova(smell_null, smell_full) # compare the null and full models
Analysis of Variance Table

Model 1: olf ~ 1
Model 2: olf ~ agecat
  Res.Df    RSS Df Sum of Sq      F    Pr(>F)    
1    179 7.7585                                  
2    175 5.6197  4    2.1388 16.651 1.395e-11 ***
---
Signif. codes:  0 ‘***’ 0.001 ‘**’ 0.01 ‘*’ 0.05 ‘.’ 0.1 ‘ ’ 1
> summary(smell_null)$sigma^2
0.04334349
> summary(smell_full)$sigma^2
0.03211259
\end{example}

This approach, which assumes all levels of \texttt{agecat} have equal error variance, favors the model with a 4 degree of freedom \texttt{agecat} effect. Note we obtain maximum likelihood estimates for the error variance of  $\hat{\sigma}^2_{\text{full}}=0.03211$. Based on Figure \ref{figure:smell}, we suspect this value may overestimate the error variance for levels 1, 2, and 3, while underestimating that of levels 4 and 5. We also suspect that the full model may be overly complex, as the means for levels 1, 2, and 3 appear to be plausibly equivalent. That is, the apparent latent grouping scheme for both regression effects and error variances is \{1,2,3\}\{4,5\}, or equivalently, \{4,5\}\{1,2,3\}. 

Next, consider the \textbf{slgf} approach. We will consider the classes of models with group-based means, group-based variances, and both group-based means and variances. We specify \texttt{dataf=smell} and \texttt{response="olf"}, along with \texttt{slgf\_beta="agecat"} and \texttt{slgf\_Sigma="agecat"} as the suspected latent grouping factor for both regression effects and variances. We set the minimum number of levels for a group to 1 with \texttt{min\_levels\_beta=1} and \texttt{min\_levels\_Sigma=1}. Note that fewer grouping schemes would be considered if we let these arguments equal 2. For simplicity, since the mean and variance grouping schemes both visually appear to be \{1,2,3\}\{4,5\}, we will restrict the schemes to be equivalent with \texttt{same\_scheme=TRUE}. Via the \texttt{usermodels} argument, we will consider the null model class \texttt{olf}$\sim$\texttt{1}, the full model class \texttt{olf}$\sim$\texttt{agecat}, and the group-means model class \texttt{olf}$\sim$\texttt{group}, which will automatically consider all possible grouping schemes. Similarly, we will consider each of these formulations with the class of both homoscedastic and group-based variances via the argument \texttt{het=c(1,1,1)}. With a relatively large amount of data, we will use the uninformative \texttt{prior="flat"}. Finally we specify a minimal training sample size of \texttt{m0=9}, although if we specify this value to be too small, \texttt{ms\_slgf()} will automatically increase it to the smallest value for which the relevant integrals converge and/or the necessary optimizations can be performed. We run \texttt{ms\_slgf} to obtain the posterior model probabilties for all 62 models under consideration. We inspect the two most probable models, with indices 62 and 32, which comprise over 99\% of the posterior probability over the model space considered: 

\begin{example}
> smell_out <- ms_slgf(dataf=smell, response="olf", lgf_beta="agecat", 
                       min_levels_beta=1, lgf_Sigma="agecat", 
                       min_levels_Sigma=1, same_scheme=TRUE, 
                       usermodels=list("olf~1", "olf~agecat", "olf~group"), 
                       het=c(1,1,1), prior="flat", m0=9)
> smell_out$models[c(1,2),c(1,2,3,5,7)]
        Model  Scheme.beta Scheme.Sigma  FModProb Cumulative
62  olf~group {4,5}{1,2,3} {4,5}{1,2,3} 0.6054935  0.6054935
32 olf~agecat         None {4,5}{1,2,3} 0.3878754  0.9933688
\end{example}

The most probable model, as suspected, is \texttt{olf}$\sim$\texttt{group}, indicating group-based means where \texttt{Scheme.beta} is \{4,5\}\{1,2,3\}. Note also \texttt{Scheme.Sigma} indicates group-based heteroscedasticity with the same scheme. This model received posterior probability of approximately 61\%. The next most probable model also has group-based heteroscedasticity with scheme \{4,5\}\{1,2,3\}, but note the model is \texttt{olf}$\sim$\texttt{agecat}, containing the full model not with group-based mean effects, but rather 4 degrees of freedom for the \texttt{agecat} effect. By inspecting \texttt{smell\_out\$scheme\_probabilities\_Sigma}, we see that models with variance grouping scheme \{4,5\}\{1,2,3\} comprise over 99\% of the posterior probability. By contrast, the models with fixed effect grouping scheme \{4,5\}\{1,2,3\} (that is, the homoscedastic and heteroscedastic versions) comprise 61\% of the posterior probability. We find these posterior probabilities intuitive, easy to interpret quantifications of uncertainty. 

The output fields \texttt{coefficients} and \texttt{variances} contain lists with the coefficients and variance(s) associated with each model. The output field \texttt{model\_fits} contains the output from a linear model fit to the model specification in question, containing the , and Note the most probable model has index \texttt{62}, so we inspect the 62nd elements of the coefficient and variance lists \texttt{smell\_out\$coefficients} and \texttt{smell\_out\$variances}, which contain the MAP estimates for each model's regression effects and variance(s), respectively. The group-based variance estimates are $\hat{\sigma}^2_{\text{\{4,5\}}}=0.0587$ and $\hat{\sigma}^2_{\text{\{1,2,3\}}}=0.0121$. We contrast these variances against the estimate $\hat{\sigma}^2_{\text{full}}=0.032$, which appears to have overestimated the variance of levels 1, 2, and 3, while simultaneously underestimating that of levels 4 and 5. 

\begin{example}
> smell_out$coefficients[[62]]
(Intercept)  group{4,5} 
  1.3252211  -0.1940328
> smell_out$variances[[62]]
     {4,5}    {1,2,3} 
0.05868885 0.01211084
\end{example}

\subsection{Case study 2: textile data}\hypertarget{subsection:textile}{}

\begin{figure}[htbp]
  \centering
  \includegraphics[width=5in]{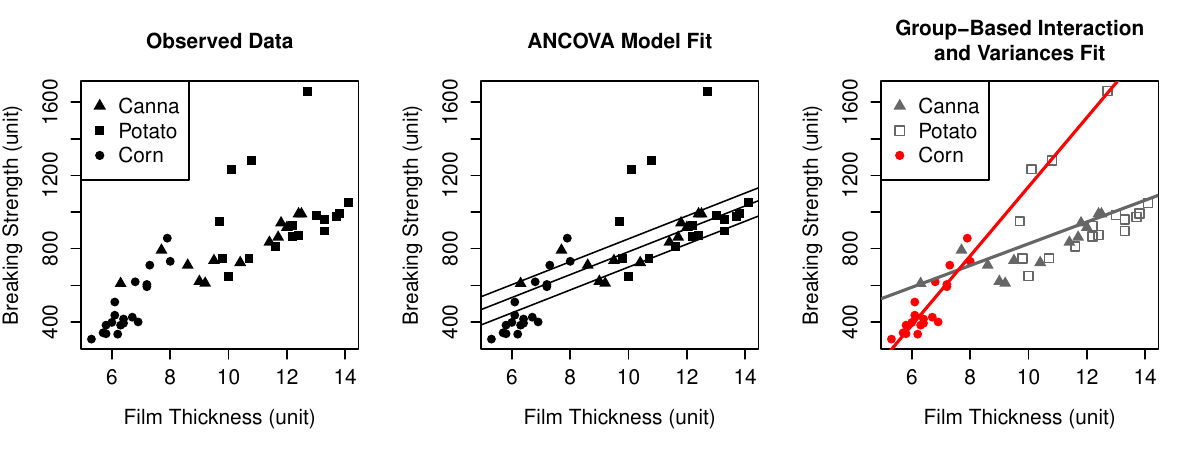}
  \caption{The breaking strength data set from \citet{Furry} represents the breaking strength of starch films depending on the thickness of a film coating and the type of starch used to make the film. The left panel shows the data. The center panel shows an ANCOVA model. The right panel shows the most probable model ($P(m|\boldsymbol{Y})\approx 66\%$) containing a latent group-based interaction between groups \{canna, potato\}\{corn\} (gray points vs. red points) and film thickness, as well as distinct variances between groups \{canna, corn\}\{potato\} (filled points vs. open points).}
  \label{figure:chips}
\end{figure}

We reanalyze the breaking strength data set of \citet{Furry}, also investigated by \citet{technometrics_paper}, to illustrate the additional flexibility of \textbf{slgf} beyond the original work. The breaking strength of a starch film \texttt{strength} (measured in grams) is analyzed according to the thickness of the film, denoted \texttt{film} (measured in $10^{-4}$ inches), and the type of starch \texttt{starch} used to create the film (canna, corn, or potato). As usual, we begin by plotting the data to ascertain whether there is a latent grouping factor present. By inspection we note that the potato films, represented by squares in Figure \ref{figure:chips}, appear to have a higher variability than the corn (filled red circles) and canna (filled gray triangles) films. 

We first illustrate a typical ANCOVA approach, in which three parallel lines for each level of starch are fit with a common error variance. This model leads to the fit shown in the center panel of Figure \ref{figure:chips}. Note only the film thickness effect is statistically significant according to a traditional hypothesis testing approach with $\alpha=0.05$. The residual standard error of this model is $\hat{\sigma}^2_{\text{ANCOVA}}=27126.09$. 

\begin{example}

> textile_ancova <- lm(strength~film+starch, data=textile)
> summary(textile_ancova)

Call:
lm(formula = strength ~ film + starch, data = textile)

Residuals:
    Min      1Q  Median      3Q     Max 
-203.63  -99.45  -57.84   56.72  637.61 

Coefficients:
             Estimate Std. Error t value Pr(>|t|)    
(Intercept)    158.26     179.78   0.880 0.383360    
film            62.50      17.06   3.664 0.000653 ***
starchcorn     -83.67      86.10  -0.972 0.336351    
starchpotato    70.36      67.78   1.038 0.304795 
\end{example}  

We contrast these findings against our methodology with \textbf{slgf}. The following arguments are input: \texttt{dataf=textile} specifies the data frame; \texttt{response="strength"} specifies the column of \texttt{textile} that contains the response variable; \texttt{slgf\_beta="starch"} and \texttt{slgf\_Sigma="starch"} indicate that the categorical variable \texttt{starch} should be used as the latent grouping factor for both regression effects and variances; \texttt{same\_scheme=FALSE} indicates that the latent regression effect and variance grouping structures do not need to be partitioned by the same levels of \texttt{starch}; \texttt{min\_levels\_beta=1} and \texttt{min\_levels\_Sigma=1} indicate that a latent group can contain only one level of \texttt{starch}; the usermodels argument indicates that we will consider main effects models \texttt{strength}$\sim$\texttt{film+starch} and \texttt{strength}$\sim$\texttt{film+starch+film*starch}, and models with group-based regression effects including \\ \texttt{strength}$\sim$\texttt{film+group} and \texttt{strength}$\sim$\texttt{film+group+film*group}; the argument \texttt{het=c(1,1,1,1)} indicates that each of these four model specifications will also be considered with group-based variances; \texttt{prior="flat"} places a flat prior on the regression effects; and \texttt{m0=8} specifies the minimal training sample size. 

\begin{example}
> data(textile)
> out_textile <- ms_slgf(dataf = textile, response = "strength", 
               lgf_beta = "starch", lgf_Sigma = "starch",
               same_scheme=FALSE, min_levels_beta=1, min_levels_Sigma=1, 
               usermodels = list("strength~film+starch", "strength~film*starch",
                                 "strength~film+group", "strength~film*group"), 
               het=c(1,1,1,1), prior="flat", m0=8)
> out_textile$models[1:5,c(1,2,3,5)]
                  Model          Scheme.beta         Scheme.Sigma      FModProb   
31  strength~film*group {corn}{canna,potato} {potato}{canna,corn}  0.6596667376 
8  strength~film*starch                 None {potato}{canna,corn}  0.3337588991 
30  strength~film*group {canna}{corn,potato} {potato}{canna,corn}  0.0018692078 
28  strength~film*group {corn}{canna,potato} {corn}{canna,potato}  0.0010854755
7  strength~film*starch                 None {corn}{canna,potato}  0.0006831597 
\end{example}

Refer to code and output above, where we provide the five most probable models. Note the three most probable models all have the latent variance grouping scheme \{potato\}\{canna, corn\}; again over 99\% of the posterior model probability is accounted for by this variance scheme. This visually agrees with the plot, which shows that the potato starch films seem to have higher variability than the canna and corn starch films. The regression effect structure is less clear: the most probable model selects the \texttt{film*group} model, which contains main effects for \texttt{film} and \texttt{group} as well as their interaction, with scheme \{canna\}\{corn, potato\}. We plot this model in the right panel of Figure \ref{figure:chips} to illustrate its plausibility. It does appear that the slope for corn is steeper than that of potato and canna, which can be contracted into a single level to simplify the model. However, the error variance for potato appears larger than that of canna and potato, as evidenced by the large spread of square potato points around the gray line. Thus we assert that the most probable model under our methodology is reasonable and appropriate. The group standard errors are $\sigma^2_{\text{\{potato\}}}=57734.046$ and $\sigma^2_{\text{\{canna,corn\}}}=5791.713$, indicating the standard ANCOVA model underestimates the error variance of the potato observations, and overestimates those of the canna and corn observations. 

Finally we illustrate the output \texttt{scheme\_probabilities\_beta} and \texttt{scheme\_probabilities\_Sigma}, which sum up the probabilities for all model specifications associated with each possible grouping scheme. We see moderately high cumulative probability for models with regression grouping scheme \{corn\}\{canna,potato\}, followed closely be models with no grouping scheme for regression effects: 

\begin{example}
> out_textile$scheme_probabilities_beta
           Scheme.beta  Cumulative
2 {corn}{canna,potato} 0.592860983
4                 None 0.403632744
1 {canna}{corn,potato} 0.002502435
3 {potato}{canna,corn} 0.001003838
\end{example}

\noindent Intuitively, based on the wider spread of the square potato points in Figure \ref{figure:chips}, we see high cumulative probability for the variance grouping scheme \{potato\}\{canna,corn\}: 

\begin{example}
> out_textile$scheme_probabilities_Sigma
          Scheme.Sigma   Cumulative
3 {potato}{canna,corn} 9.975853e-01
2 {corn}{canna,potato} 2.184257e-03
1 {canna}{corn,potato} 2.304323e-04
4                 None 1.632320e-08
\end{example}
\subsection{Case study 3: locknut data}\hypertarget{subsection:torque}{}

We consider the two-way replicated layout of \citet{locknut}, where the torque (\texttt{torque}) required to tighten a locknut was measured as a function of a plating process (\texttt{plating}) and a threading method (\texttt{fixture}). 

\begin{figure}[htbp]
  \centering
  \includegraphics[]{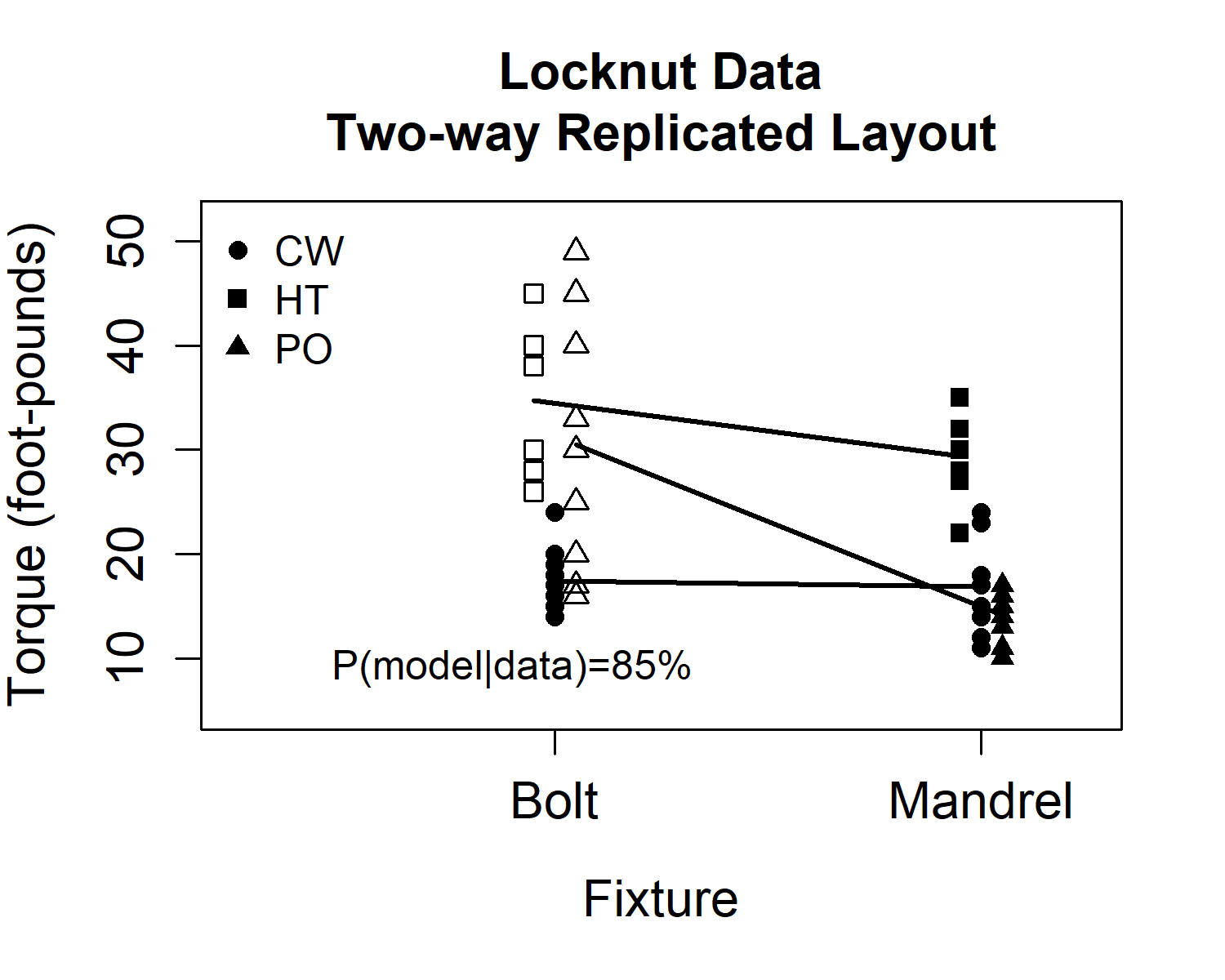}
  \caption{The most probable model ($P(m|Y)\approx 85\%$) contains a full fixture by plating interaction effect with no grouping structure, and group-based variances based on the levels of this interaction with scheme \{bolt*CW, mandrel*CW, mandrel*HT, mandrel*PO\}\{bolt*HT, bolt*PO\} (filled points vs. open points).}
  \label{figure:locknut}
\end{figure}

A two-way analysis with an interaction yields the following ANOVA table. The fixture and plating main effects, along with fixture by plating interaction, are all statistically significant at level $\alpha=0.005$. Additionally, we find $\hat{\sigma}^2_{\text{Full}}=36.58$: 

\begin{example}
> anova(lm(Torque~Fixture+Plating+Fixture*Plating, data=locknut))  
Analysis of Variance Table  

Response: Torque
                Df Sum Sq Mean Sq F value    Pr(>F)    
Fixture          1  821.4  821.40 22.4563 1.604e-05 ***
Plating          2 2290.6 1145.32 31.3118 9.363e-10 ***
Fixture:Plating  2  665.1  332.55  9.0916 0.0003952 ***
Residuals       54 1975.2   36.58  
\end{example}

Upon inspection of Figure \ref{figure:locknut}, we suspect that two latent characteristics are at play. First, based on the non-parallel lines representing the plating effects, there may be a group-by-plating interaction, so we will consider \texttt{slgf\_beta="Plating"}. Note since fixture has only two levels, it is not feasible to consider group-based effects based on fixture since the one degree of freedom fixture effect would be equivalent to a group effect. 

Regarding the variance structure, the variance of the torque amount at levels PO and HT appears higher, but only for the bolt fixture. This suggests that the levels of the interaction govern the variance groups; that is, \texttt{slgf\_Sigma="Fixture*Plating"}. Since this specific variable header does not appear in the \texttt{locknut} data set, we  manually create a new variable with each interaction level by pasting together the main effect variables: 

\noindent \texttt{locknut\$Interaction <- paste0(locknut\$Fixture, "*", locknut\$Plating)}

Thus we consider the following model specifications. \citet{Liangetal} (p. 420) note that the Zellner-Siow mixture of $g$-prior provides a fully Bayesian, consistent model selection procedure for small $n$ along with relatively straightforward expressions for the marginal model probabilities. This approach is implemented by the user with the argument \texttt{prior="zs"}: 

\begin{example}
> data(locknut)
> locknut$Interaction <- paste0(locknut$Fixture, "*", locknut$Plating)
> out_locknut <- ms_slgf(dataf=locknut, response="Torque", same_scheme=FALSE, 
                       lgf_beta="Plating", min_levels_beta=1,
                       lgf_Sigma="Interaction", min_levels_Sigma=1, 
                       usermodels=list("Torque~Fixture+Plating+Fixture*Plating", 
                                       "Torque~Fixture+group+Fixture*group"), 
                       het=c(1,1), prior="zs", m0=2)
\end{example}

This formulation favors the same main and interaction effects favors by the standard model. However, \textbf{slgf} favors group-based variances with scheme \{bolt*HT, bolt*PO\}\{bolt*CW, mandrel*CW, mandrel*HT, mandrel*PO\} with posterior probability of approximately 85\%. This variance structure was expected based on the relatively larger spread of the open points in Figure \ref{figure:locknut}. As we have noted previously, the group variance estimates show that the heteroscedastic model overestimates the variance for some levels of fixture and plating, and underestimates it for others. Since model `13` was the model probable model, we print these variances, obtaining $\hat{\sigma}^2_{\text{bolt*HT,bolt*PO}}\approx 85.0$ and $\hat{\sigma}^2_{\text{bolt*CW,mandrel*CW,mandrel*HT,mandrel*PO}}\approx 11.6$: 

\begin{example}
> out_locknut$variances[[13]]
{bolt*HT,bolt*PO} {bolt*CW,mandrel*CW,mandrel*HT,mandrel*PO} 
         85.00448                                   11.58652
\end{example}

\subsection{Case study 4: bottles data}\hypertarget{subsection:bottles}{}

\begin{figure}[htp!]
  \centering
  \includegraphics[width=5in]{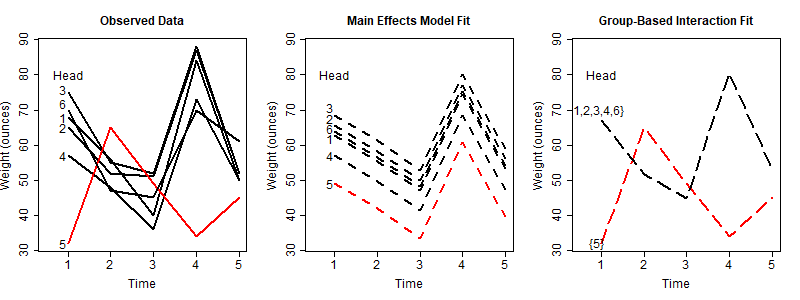}
  \caption{The bottles data set from \citet{bottles} represents fill weights by six machine heads over five time points. The left panel shows the data, with head 5 appearing to be out of calibration. The center panel shows a main effects model, with a realistic fit for heads 1, 2, 3, 4, and 6, but not 5. The right panel shows the most probable group-based interaction ($P(m|\boldsymbol{Y})> 99.9\%$) with main effects for time and a group-by-time interaction with scheme \{5\}\{1,2,3,4,6\}.}
  \label{figure:bottles}
\end{figure}

Finally, we consider the data of \citet{bottles}, where a machine with six heads (\texttt{head}) is designed to fill bottles (\texttt{weight}). The weight of each bottle is measured once over five time points (\texttt{time}) as a two-way unreplicated layout. A visual inspection of the data (Figure \ref{figure:bottles}, left panel) indicates that one of the filling heads is behaving distinctly than the other five. There appears to be an interaction between head and time, but we lack the degrees of freedom to fit such a model. If we were to fit the standard main effects model, we obtain the clearly inappropriate model fit in the center panel of Figure \ref{figure:bottles}. 

Since it appears that head \{5\} is out of calibration in some way as compared to heads \{1,2,3,4,6\}, we instead consider the group-based interaction model \texttt{weight}$\sim$\texttt{time+group:time} where `head' is the regression effect SLGF. For this illustration, we consider only homoscedastic models. In this data-poor context, we recommend the use of the Zellner-Siow mixture of $g$-prior by specifying \texttt{prior="zs"} in the \texttt{ms\_slgf} function. The minimal training sample size can be much lower, as this prior is proper. We inspect the posterior model probabilities of the most probable model and the additive main effects model: 

\begin{example}
> bottles_me <- lm(weight~time+heads, data=bottles)
> bottles2 <- data.frame(weight=bottles$weight, time=as.factor(bottles$time),
                         heads=as.factor(bottles$heads))
> bottles_out <- ms_slgf(dataf=bottles2, response="weight", lgf_beta="heads", 
                 min_levels_beta=1, lgf_Sigma=NA, min_levels_Sigma=NA, same_scheme=FALSE, 
                 usermodels=list("weight~time+group:time",  "weight~time+heads"),
                 het=c(0,0), prior="zs", m0=2)
> bottles_out$models[1:2,c(1,2,4,5)]
                   Model    Scheme.beta Log-Marginal  FModProb
5 weight~time+group:time {5}{1,2,3,4,6}     -103.168    0.9991932
32     weight~heads+time           None     -114.726 0.0002158313 

\end{example}

The group-based approach overwhelmingly favors the model with a main effect for `time` along with the group-based interaction `group:time` with scheme \{5\}\{1,2,3,4,6\}. We also note that the error variance for the main effects model is $\hat{\sigma}^2_{\text{ME}}=130.1233$, while the estimate for the group-based interaction model is $\hat{\sigma}^2_{\text{\{5\}\{1,2,3,4,6\}}}=39.76$, suggesting the main effects model seriously overestimates the error variance and thus may lead to misleading inference on regression effects. 

We note that there will be a linear dependency between the group-by-time interaction and the time main effect for time 5. The \texttt{NA} values can be seen by inspecting the coefficients of the corresponding model. These effects are not counted in the dimensionality of the model when computing $q^b(\boldsymbol{Y}|m)$. 

\begin{example}
> bottles_out3$coefficients[[5]]
           (Intercept)            heads2                    heads3            heads4 
                 53.24              1.80                      4.80             -6.80 
                heads5            heads6    group{1,2,3,4,6}:time1    group{5}:time1 
                 -8.24             -1.00                     14.00            -13.00 
group{1,2,3,4,6}:time2    group{5}:time2    group{1,2,3,4,6}:time3    group{5}:time3 
                 -1.40             20.00                     -8.20              4.00 
group{1,2,3,4,6}:time4    group{5}:time4    group{1,2,3,4,6}:time5    group{5}:time5 
                 27.40            -11.00                        NA                NA 
\end{example}

\section{Conclusion} \hypertarget{section:conclusion}{}

This manuscript has provided an overview of the \textbf{slgf} package in R, which is available from the Comprehensive R Archive Network. Source code can be found on Github at \url{https://github.com/metzger181osu/slgf}. The \textbf{slgf} package allows the user to determine whether latent groupings of categorical predictor's levels provide a better characterization of the response variable compared with ordinary linear models that do not account for the suspected latent groupings. This is accomplished through the \textit{suspected latent grouping factor} methodology of \cite{technometrics_paper}. The methodology allows for formal comparisons between ordinary linear models and latent grouping models, which protects the user from automatically selecting a spurious clustering structure that is not well supported by the data. We illustrate the ability to detect the lack of a grouping structure in the  simulation studies of \citet{technometrics_paper}. 

The \textbf{slgf} package allows the user to (i) explore different grouping schemes for fixed effects and error variances, and (ii) specify entirely separate latent grouping factors for fixed effects and variances. We illustrate (i) in {Case Study 2: Textile data}, where the top model shows a different regression line for corn compared to canna and potato, but the error variance for potato is different from canna and corn (see Figure \ref{figure:chips}). To show (ii), we considered the locknut example of Subsection \hyperlink{subsection:torque}{\textbf{Case study 3: locknut data}}, where we considered whether fixture (bolt, mandrel) exhibited a fixed effect latent grouping structure, and whether interaction (bolt*CW, bolt*HT, bolt*PO, mandrel*CW, mandrel*HT, mandrel*PO) exhibited a variance latent grouping structure. As described in Subsection \hyperlink{subsection:torque}{\textbf{Case study 3: locknut data}}, we found no latent grouping structure for fixed effects, but torque error variance for bolt*HT and bolt*PO differ from the other interaction levels. The analysis supported no latent grouping structure for plating. 

The \textbf{slgf} package provides functionality to detect plausible underlying cluster structures among levels of categorical predictors in linear models. This exercise in cluster detection is in some ways similar to considering a finite mixture model. R packages already exist to fit finite mixture models using the EM algorithm, such as \textbf{mixtools} \citep{mixtools}. The \textbf{flexmix} package \citep{flexmix} in particular is notable for its ability to fit mixture models to regression data (including Gaussian, binomial, and Poisson models). Additionally, the package \textbf{MultiLCIRT} also considers latent variables for the item response theory setting; see \citet{BARTOLUCCI2014971}, who use BIC for model selection rather than fractional Bayes factors. 

In contrast to fitting finite mixture models for the purpose of parameter estimation and inference, \textbf{slgf} assesses the plausibility of cluster structures for small to medium-sized data sets via model selection. Additionally, \textbf{slgf} can avoid problems with convergence in the EM algorithm that may arise in small-sample scenarios, particularly when the number of data points is relatively low and the model being fit (e.g., a two component mixture model) is larger than the actual model generating the data (e.g., a one component mixture model with no cluster structure).

By contrast, \textbf{slgf} circumvents convergence issues by considering all possible groupings of points within the user-specified model classes, obtaining integrated likelihoods and posterior model probabilities for each model, and quantifying overall probability of a  cluster structure as the sum of all posterior probabilities for models with two groups by the law of total probability. The \textbf{slgf} package thus excels in smaller-data settings where assessing the plausibility of a cluster structure is the core goal, and packages like \textbf{flexmix} will excel in cases where the main goal is to fit specified mixture models and conduct inference on parameters.

In addition to the basic \textbf{slgf} demonstration shown in {Case Study 1: Smell data}, we illustrate \textbf{slgf} functionality for mixtures of $g$-priors \citep{Liangetal} and a two way unreplicated layout in Case Study 4: Bottles data. Mixtures of $g$-priors have been shown to work well with fractional Bayes factor methods to reduce the training fraction when sample size is small relative to the number of model parameters \citep{technometrics_paper}. 

Finally, although the methodology described here and in \citet{technometrics_paper} exclusively handles two latent groups, we call on any readers with a compelling data set that may exhibit more than two latent groups to contact the authors so that we might explore a generalization of our method to more than two groups.  

We have provided an overview of functionality that we hope will enable scientists from diverse fields to access the SLGF methodology of \cite{technometrics_paper} via the \textbf{slgf} package to detect hidden groupings in the levels of categorical predictors that might impact outcomes of interest across a wide range of human endeavors.  

\newpage 

\bibliography{references}



\end{document}